# Mathematical modeling of regime-dependent thermal response in multilayer wall assemblies with equal total steady thermal resistance

**Dmytro Usenko[1], Valery Usenko[2], Iryna Usenko[3]**
[1]Department of Higher and Applied Mathematics and Physics, National University, Yuri Kondratyuk Poltava Polytechnic, Poltava, Ukraine
[2, 3]Department of Construction and Civil Engineering, National University "Yuri Kondratyuk Poltava Polytechnic", Poltava, Ukraine
[1]Corresponding author
**E-mail:** [1]*usenko_phd@ukr.net*, [2]*valery_usenko@ukr.net*, [3]*irina__usenko@ukr.net*

**Abstract.** Steady thermal resistance is a fundamental characteristic of building-envelope assemblies; however, equal steady resistance does not uniquely determine their transient thermal behavior. This study aims to numerically separate the effects of layer order, temperature-dependent thermophysical properties, and redistribution of thermal resistance between bulk and interfacial components in stationarily equivalent multilayer wall assemblies subjected to periodic and finite-pulse thermal loading. A one-dimensional transient heat-conduction model with Robin boundary conditions, finite interfacial thermal resistance, and source-supported temperature-dependent material properties was employed. All compared configurations were constrained to the same total steady thermal resistance at the reference temperature, $R_{tot}$ = 3.65315 m$^2$ K\/W. The numerical implementation passed the adopted verification procedure with restrictions. The combined grid/time uncertainty was 0.325 %, below the prescribed 0.5 % acceptance criterion. For the tested pair differing only in layer order, the relative effect was –0.905 % for the periodic decrement factor and +0.566 % for the finite-pulse peak transmission ratio, with effect-to-uncertainty ratios of 79.2 and 88.9, respectively; the opposite signs therefore refer to two distinct regime-specific metrics. The temperature-dependent configuration produced a numerically resolved periodic effect of –2.005 %, whereas the corresponding pulse effect of –0.282 % remained below the adopted numerical-resolution threshold. For a compensated 5-10 % redistribution of bulk thermal resistance to interfacial contacts, the additional dynamic effect was likewise unresolved. The results demonstrate the value of controlled factor separation combined with explicit numerical-uncertainty assessment when evaluating transient thermal behavior.

**Keywords:** multilayer wall assembly, transient heat conduction, steady-resistance equivalence, layer order, temperature-dependent thermophysical properties, interfacial thermal resistance, thermal pulse.

## 1. Introduction

Steady thermal resistance and thermal transmittance are fundamental descriptors of building-envelope performance but do not fully characterize transient behavior. Assemblies with identical or similar steady thermal resistance may exhibit different responses because of the spatial distribution of thermal resistance and heat-storage capacity [1, 2].

The influence of internal structure and layer arrangement on periodic thermal characteristics is established [1, 3] and is commonly described using the decrement factor, phase lag, and other non-interchangeable dynamic metrics [4]. Accordingly, the layer-order effect itself is treated here as background rather than as a standalone contribution.

Additional effects arise from temperature-dependent thermophysical properties and imperfect thermal contact between adjacent layers. Experimentally derived $k(T)$ and $cp(T)$ relations demonstrate that these dependencies may need to be considered within their validated temperature range [5], while nonlinear transient heat-conduction problems with temperature-dependent material properties have been independently investigated and numerically benchmarked [6]. Finite interfacial thermal resistance, in turn, permits a temperature discontinuity across an interface while maintaining heat-flux continuity [7, 8].

Periodic excitation is naturally characterized by harmonic amplitude and phase quantities, whereas a finite thermal disturbance requires metrics associated with peak transmission and the temporal structure of the transient response. Recent pulse-based methods demonstrate a mathematical relationship between finite-pulse and periodic thermal characterization [9]; nevertheless, this relationship does not predetermine whether the sign or relative magnitude of the influence of a particular structural factor will remain unchanged when the excitation regime changes. Moreover, direct transfer of dynamic metrics between different experimental and physical settings requires caution, especially where additional transport mechanisms are involved [10].

The methodological gap lies in the insufficiently controlled separation of layer order, temperature-dependent properties, and interfacial resistance within a single stationarily equivalent system, because previous studies often vary several of these factors simultaneously.

This study assesses these factors separately under equal total steady thermal resistance and explicit numerical-uncertainty control. The novelty lies in their controlled cross-regime separation and assessment; for the tested layer-order pair, the regime-specific principal metrics exhibit resolved effects of opposite sign.

## 2. Mathematical model and methodology

### 2.1. Mathematical formulation

The three-layer assembly comprises: $A$ = 0.10 m LECA lightweight concrete, $k$ = 1.1665 W/m·K, $\rho$ = 1831 kg/m$^3$, $c_p$ = 868.452 J/kg·K [5]; $B$ = 0.12 m insulation, $k$ = 0.038 W/m·K, $\rho$ = 32 kg/m$^3$, $c_p$ = 840 J/kg·K [2]; and $C$ = 0.12 m brick, $k$ = 0.72 W/m·K, $\rho$ = 1920 kg/m$^3$, $c_p$ = 835 J/kg·K [2], all values referring to $T_{ref}$ = 20 °C. The external/internal heat-transfer coefficients are 10/7 W/(m$^2$·K), and the indoor air temperature is 20 °C.

For layer $i$, conservation of energy is expressed as:

$$\rho_i(T)c_{p,i}(T)\frac{\partial T_i}{\partial t} = \frac{\partial}{\partial x}\left[k_i(T)\frac{\partial T_i}{\partial x}\right], \tag{1}$$

where $T_i$ is temperature, °C; $t$ is time, s; $x$ is the spatial coordinate, m; $k_i(T)$ is thermal conductivity, W/(m·K); $c_{p,i}(T)$ is specific heat capacity, J/(kg·K); and $\rho_i(T)$ is density, kg/m$^3$.

The local heat flux is given by Fourier's law:

$$q_i = -k_i(T)\frac{\partial T_i}{\partial x}. \tag{2}$$

Robin boundary conditions are imposed on the external and internal surfaces:

$$q_{out} = h_{out}\left[T_e(t) - T_{s,out}(t)\right], \quad q_{in} = h_{in}\left[T_{s,in}(t) - T_i\right]. \tag{3}$$

For an imperfect interface, heat-flux continuity is retained while a finite temperature jump is permitted: $q_j^- = q_j^+ = q_j$, $T_j^- - T_j^+ = q_j R_{(c,j)}$. Here $R_{c,j}$ is the interfacial thermal resistance, m$^2$·K/W.

In configuration $N$, only layer $A$ is temperature dependent. The experimentally fitted relations

[5] are $k_A(T)$ = 1.1845-0.0009 T W/(m·K) and $c_{p,A}(T)$ = 818.14+2.5156T J/(kg·K), where $T$ is in °C, $0 \le T \le 50$°C; and $\rho_A = 1831$ kg/m$^3$ remains constant. These relations are used only for the source-supported oven-dry $LECA$ control mixture and are not extrapolated to layers B or C.

## 2.2. Controlled configurations

The total steady thermal resistance at the reference temperature is defined as $R_{tot} = \frac{1}{h_{out}} + \sum_i \frac{L_i}{k_i(T_{ref})} + \sum_j R_{c,j} + \frac{1}{h_{in}}$, where $L_i$ is layer thickness and $T_{ref}$ is the reference temperature.

For configurations B, O, N, C5, and C10, $R_{tot}$ = 3.65315 m$^2$ K/W at the reference temperature.

**Table 1.** Controlled configurations and steady-resistance equivalence of the multilayer system

| Config. | Layer order | Property model | $R_c$ share, % | $R_{bulk}$, m$^2$K/W | $R_c$, m$^2$K/W | $R_{tot}$, m$^2$K/W | $C_A$, kJ/(m$^2$K) |
|---|---|---|---|---|---|---|---|
| B | A-B-C | Constant | 0 | 3.4103 | 0 | 3.65315 | 354.6 |
| O | C-B-A | Constant | 0 | 3.4103 | 0 | 3.65315 | 354.6 |
| N | A-B-C | Temperature-dependent | 0 | 3.4103 | 0 | 3.65315 | 354.6 |
| C5 | A-B-C | Compensated constant | 5 | 3.2398 | 0.1705 | 3.65315 | 354.6 |
| C10 | A-B-C | Compensated constant | 10 | 3.0693 | 0.341 | 3.65315 | 354.6 |

B denotes the baseline A-B-C sequence; O uses the reversed C-B-A sequence, whereas $N$ retains A-B-C and activates the temperature-dependent properties of layer $A$.

In C5 and C10, 5 % and 10 % of the baseline $R_{bulk}$, respectively, are assigned to interfacial thermal resistance and divided equally between the A-B and B-C interfaces. Thus, each interface has $R_c \approx 0.0853$ m$^2$ K/W in C5 and $R_c \approx 0.1705$ m$^2$ K/W in C10. The conductivity of layer B is then scaled, without changing its thickness, to preserve $R_{tot}$. These cases therefore represent redistribution between bulk and interfacial resistance rather than isolated addition of contact resistance. $C_A$ in Table 1 is a reference-temperature value and is not constant during transient $N$ calculations.

## 2.3. Thermal loading and response metrics

For periodic loading, $T_e(t) = \bar{T}_e + A_e \sin[2\pi(t - t_\phi)/P]$, with $P$ = 24 h, $\bar{T}_e$ = 30 °C, and $A_e$ = 10 K. The response metrics are the decrement factor $f_T = A_{s,in}/A_e$ and harmonic phase lag $\Delta t_\phi = [(\varphi_e - \varphi_{s,in}) \ mod \ 2\pi]/\omega$, $\omega = 2\pi/P$. The finite triangular pulse rises from a baseline external temperature of 20 °C by 20 K over 3 h and decays over the following 3 h; the corresponding steady thermal state is used as the initial condition.

The principal pulse-response metric is the peak transmission ratio $r_{peak} \frac{max[T_{s,in}(t) - T_{s,in,0}]}{\Delta T_{pulse}}$. Where $T_{s,in,0}$ is the reference internal-surface temperature before pulse application. The time-to-peak delay $\Delta t_{peak}$ and centroid delay $\Delta t_c$ are also evaluated. These quantities characterize a finite transient process and are not interpreted as harmonic phase lag.

## 2.4. Numerical method and verification

The equations are discretized by a conservative cell-centered finite-volume method, integrated in time using implicit Backward Euler, and updated for temperature-dependent properties by Picard iterations. These standard procedures are not treated as scientific contributions.

The baseline resolution is 40 control volumes per layer with a time step of 300 s. Additional calculations employ 80/300 s, 40/150 s, and 80/150 s combinations for spatial, temporal, and combined refinement.

The numerical implementation was assessed using three benchmark levels: B1, a linear multilayer benchmark [1]; B2, an imperfect-interface benchmark [7, 8]; and B3, a moderately nonlinear problem with temperature-dependent properties [6]. The overall verification status is PASS WITH RESTRICTIONS. For B2, published pointwise temperature data are insufficient for a complete independent reconstruction of the reference temperature history. For B3, the published agreement with the finite-element solution is below 4 %, but tabulated COMSOL ordinates are unavailable for a new pointwise comparison.

The verification procedure therefore assesses the correctness of the numerical implementation but does not constitute full experimental validation of the present model.

Grid and time-step uncertainties, $u_g$ and $u_t$, were estimated from the maximum relative changes in the periodic decrement factor and phase lag between the two finest grid-only and time-only solutions, respectively. The combined uncertainty was evaluated as $u_{grid+time} = \left(u_g^2 + u_t^2\right)^{1/2}$ = 0.325 %, below the adopted 0.5 % criterion. For N, C5, and C10, numerical resolvability was characterized by $\eta = |\Delta|/u_{grid+time}$, with $\eta > 1$ interpreted as a resolved effect. For the key B/O comparison, a dedicated four-level refinement was used: the uncertainty in the relative effect was estimated from its deviations between the combined-fine, grid-fine, and time-fine solutions, and sign stability was additionally checked across all four resolutions.

## 3. Results and discussion

Figs. 1 and 2 show time-dependent deviations from B normalized by the maximum absolute amplitude of the baseline response. These quantities should not be interpreted as the relative changes in the principal regime-specific metrics reported below.

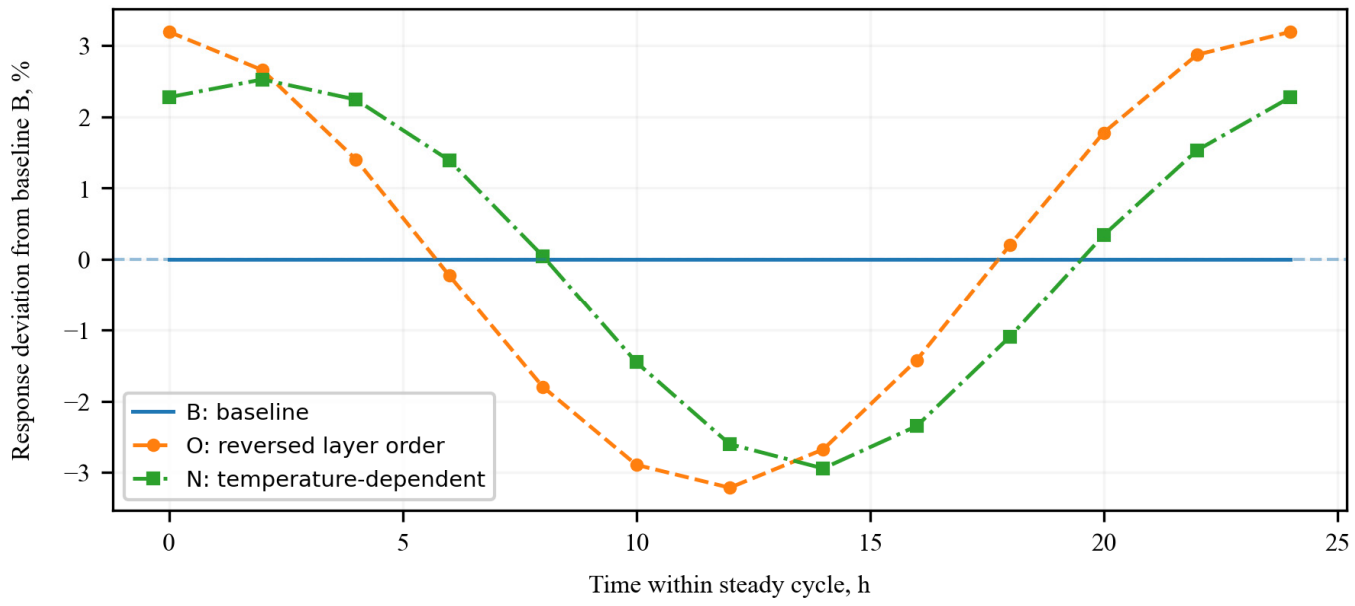

**Fig. 1.** Influence of layer order and temperature-dependent properties on the periodic thermal response relative to the baseline configuration

For O, the relative change in the principal periodic metric is $\Delta_O^{(P)}$ = –0.905 %, with an effect-to-uncertainty ratio of $|\Delta|/u =$ 79.2. The layer-order effect is therefore clearly resolved numerically for the tested configuration.

For configuration $N$, $\Delta_N^{(P)}$ = –2.005 %, and $|\Delta|/u$ = 6.17. Thus, the temperature-dependent configuration produces a numerically resolved periodic effect within the investigated temperature range.

For O, the finite-pulse effect is $\Delta_O^{(I)}$ = +0.566 %, with $|\Delta|/u$ = 88.9.

For the tested B/O pair, the layer-order effect is –0.905 % for periodic $f_T$ and +0.566 % for pulse $r_{peak}$. Because these are different regime-specific metrics, the opposite signs do not represent reversal of one common physical quantity; they indicate regime-dependent response within the adopted metric framework. Both effects remain sign-stable under refinement and are strongly resolved ($|\Delta|/u = 79.2$ and 88.9), but this result is specific to the tested pair.

For configuration $N$, the finite-pulse effect is $\Delta_N^{(I)} = -0.282$ %, with $|\Delta|/u = 0.868 < 1$.

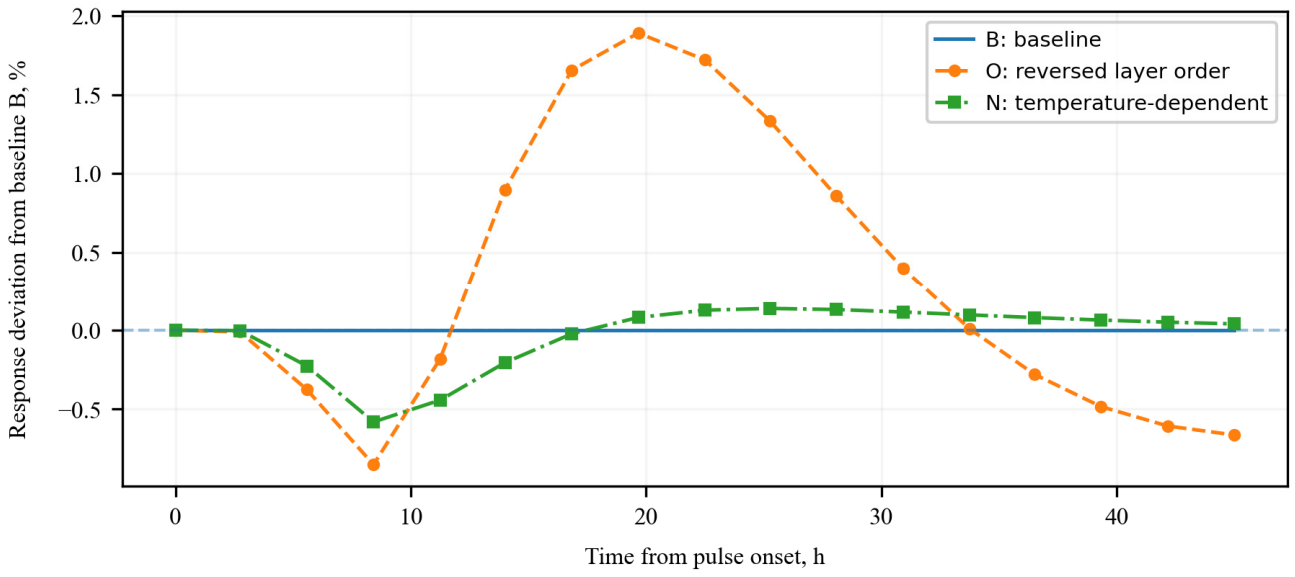

**Fig. 2.** Influence of layer order and temperature-dependent properties on the finite-pulse thermal response relative to the baseline configuration

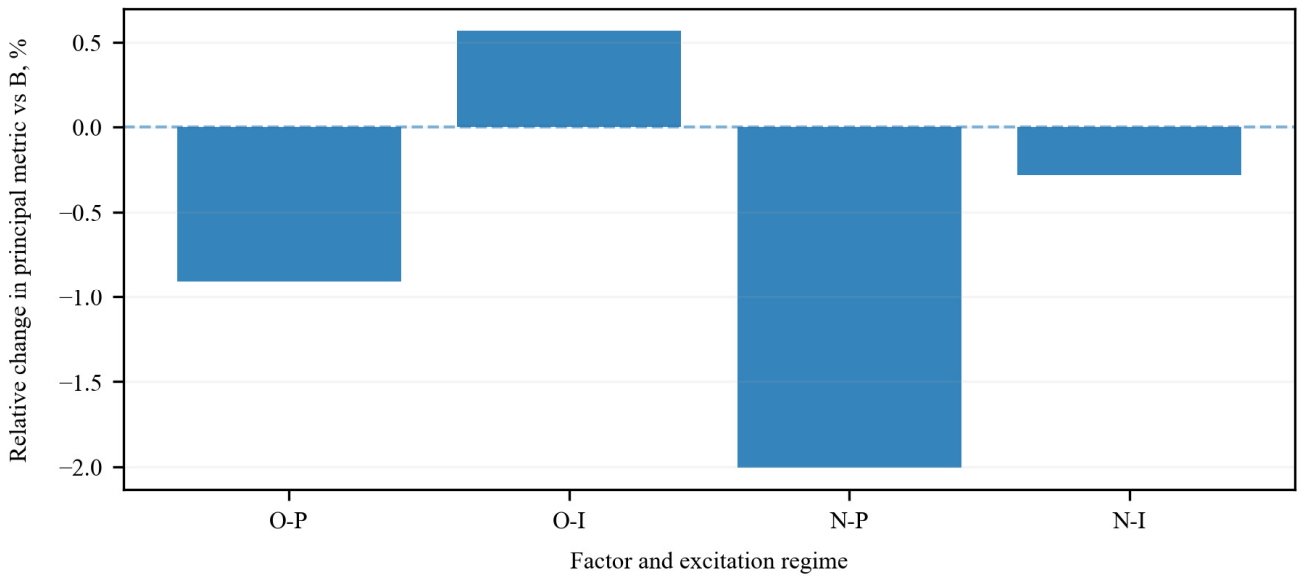

**Fig. 3.** Regime-dependent effects of layer order and temperature-dependent properties on the principal dynamic-response metrics

Accordingly, the pulse effect of the temperature-dependent configuration is not resolved relative to the adopted numerical-uncertainty budget. This result should not be interpreted as evidence that the underlying physical effect is zero.

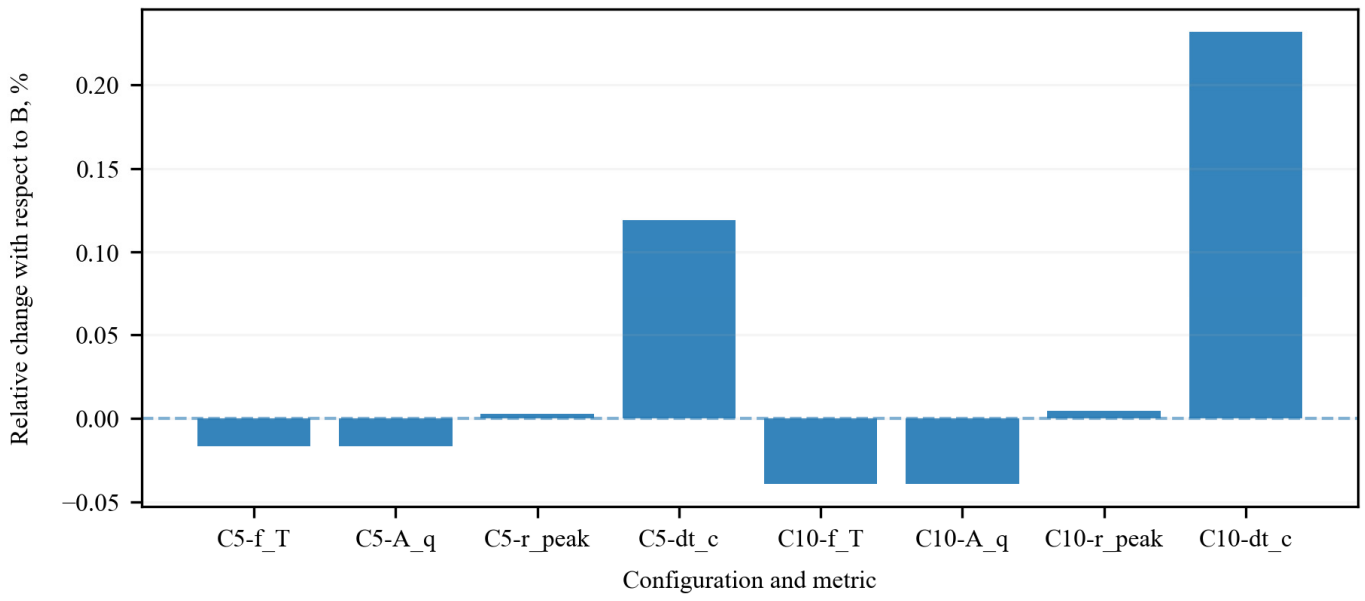

**Fig. 4.** Sensitivity of transient thermal-response metrics to compensated interfacial thermal resistance at unchanged total steady thermal resistance

For C5, the effect-to-uncertainty ratios for the principal periodic and pulse metrics are approximately 0.0515 and 0.0077, respectively. For C10, the corresponding values are 0.1215 and 0.0137.

Thus, controlled redistribution of 5-10 % of the baseline $R_{bulk}$ to interfacial thermal resistance does not produce a numerically resolved change in the principal dynamic-response metrics.

Fig. 4 illustrates nominal multi-metric sensitivity, whereas the conclusion regarding the unresolved nature of the C5/C10 effects is based on the $|\Delta|/u$, ratios reported in Table 2.

**Table 2.** Dynamic thermal-response metrics and numerical resolvability of the controlled configurations

| Config. | $f_T$, 1 | $\Delta t_{\varphi}$, h | $r_{peak}$, 1 | $\Delta t_{peak}$, h | $\Delta t_c$, h | $\lvert\Delta\rvert/u(P)$, 1 | $\lvert\Delta\rvert/u(I)$, 1 |
|---|---|---|---|---|---|---|---|
| B | 0.007548 | 10.581 | 0.005451 | 10 | 17.975 | | |
| O | 0.00748 | 10.697 | 0.005482 | 10.25 | 17.665 | 79.2 | 88.9 |
| N | 0.007397 | 10.651 | 0.005436 | 10.083 | 18.014 | 6.1656 | 0.8677 |
| C5 | 0.007547 | 10.604 | 0.005452 | 10.083 | 17.997 | 0.0515 | 0.0077 |
| C10 | 0.007545 | 10.626 | 0.005452 | 10.083 | 18.017 | 0.1215 | 0.0137 |

## 4. Conclusions

1) A controlled numerical framework was established for multilayer wall assemblies with $R_{\text{tot}} = 3.65315$ m$^2$ K/W at the reference temperature, separately assessing layer order, temperature-dependent properties, and bulk-to-interface resistance redistribution.

2) For the tested B\/O pair, the relative layer-order effect is –0.905 % for the periodic decrement factor $f_T$ and +0.566 % for the finite-pulse peak transmission ratio $r_{peak}$. The corresponding effect-to-uncertainty ratios of 79.2 and 88.9 confirm that both regime-specific effects are numerically resolved. The opposite signs concern two different regime-specific metrics and therefore indicate regime dependence rather than reversal of a single common physical quantity; the finding should not be extrapolated automatically to other multilayer assemblies.

3) The temperature-dependent configuration produced a numerically resolved periodic effect of –2.005 %, whereas the corresponding pulse effect of –0.282 % remained unresolved.

4) For compensated redistribution of 5-10 % of the baseline bulk thermal resistance to interfacial contacts, the principal dynamic effects remained well below the adopted numerical-resolution threshold. A strong additional contact-related effect is therefore not supported within the investigated range.

5) The combined grid/time numerical uncertainty was 0.325 %, below the adopted acceptance criterion of 0.5 %. The numerical implementation has a verification status of PASS WITH RESTRICTIONS; the present results should not be interpreted as full experimental validation of the model.

6) The conclusions are limited to one-dimensional dry heat conduction, the adopted material-property laws and temperature ranges, the tested B/O pair, and the C5/C10 configurations. Moisture transport, phase change, two- and three-dimensional thermal bridges, and radiative heat transfer in the main model are outside the scope of this study.

## Acknowledgements

The authors have not disclosed any funding.

## Author contributions

Dmytro Usenko: Conceptualization, Methodology, Formal Analysis, Software, Validation, Data Curation, Visualization, Writing – Original Draft Preparation, Writing – Review and Editing, Project Administration. Valery Usenko: Conceptualization, Investigation, Methodology, Formal Analysis, Validation, Resources, Writing – Review and Editing. Iryna Usenko: Investigation, Validation, Resources, Supervision, Writing – Review and Editing.

## Data availability

The datasets generated during and/or analyzed during the current study are available from the corresponding author on reasonable request.

## Conflict of interest

The authors declare that they have no conflict of interest.